\documentclass[twocolumn,showpacs,preprintnumbers,amsmath,amssymb]{revtex4}
\usepackage{graphicx}% Include figure files
\usepackage{dcolumn}% Align table columns on decimal point
\usepackage{bm}% bold math

\begin{document}

\title{Effect of dipolar moments in domain sizes of lipid bilayers and monolayers}

\author{A. Travesset}
\affiliation{Ames Laboratory and Department of Physics and
Astronomy, Iowa State University, Ames, Iowa 50011}
\date{\today}
\email{trvsst@ameslab.gov}
\date{\today}

\begin{abstract}

Lipid domains are found in systems such as multi-component bilayer
membranes and single component monolayers at the air-water
interface. It was shown by Andelman et al. (Comptes Rendus 301,
675 (1985)) and McConnell et al. (Phys. Chem. {\bf 91}, 6417
(1987)) that in monolayers, the size of the domains results from
balancing the line tension, which favors the formation of a large
single circular domain, against the electrostatic cost of
assembling the dipolar moments of the lipids. In this paper, we
present an exact analytical expression for the electric potential,
ion distribution and electrostatic free energy for different
problems consisting of three different slabs with different
dielectric constants and Debye lengths, with a circular
homogeneous dipolar density in the middle slab. From these
solutions, we extend the calculation of domain sizes for
monolayers to include the effects of finite ionic strength,
dielectric discontinuities (or image charges) and the
polarizability of the dipoles and further generalize the
calculations to account for domains in lipid bilayers. In
monolayers, the size of the domains is dependent on the different
dielectric constants but independent of ionic strength. In
asymmetric bilayers, where the inner and outer leaflets have
different dipolar densities, domains show a strong size dependence
with ionic strength, with molecular-sized domains that grow to
macroscopic phase separation with increasing ionic strength. We
discuss the implications of the results for experiments and
briefly consider their relation to other two dimensional systems
such as Wigner crystals or heteroepitaxial growth.

\end{abstract}
 \maketitle

\maketitle

\section{Introduction}

Single component monolayers of phospholipids or fatty acids at the
air-water interface exhibit stable, micron sized domains of
coexisting phases \cite{Losche1983,Peter1983,McConnell1984}. Such
domains have not been observed in single component bilayers, but
vesicles consisting of mixtures, such as ternary mixtures of
phospholipids, sphingolipids and cholesterol, which have been the
subject of intense investigations \cite{Simons2004,Veatch2005}, do
exhibit stable domains. Most of the natural occurring
phospholipids and sphingolipids are zwitterionic
\cite{Israelachvili2000} and exhibit a permanent electric dipole
moment, which is quite significant, around
$18$D\cite{Gawrisch1992}. In a monolayer, these dipoles all point
in the same direction, with its major component perpendicular to
the air-water interface. In single component bilayers, the
permanent dipoles in both leaflets point in opposite direction and
the membrane does not have a permanent dipole. In many important
situations, however, the lipid composition within the inner and
outer leaflet is asymmetric. In most cell membranes for example,
phosphatidylethanolamine (PE) and phosphatidylserine (PS) are
preferably found in the inner leaflet (in contact with the
cytoplasm) while sphingomyelin (SM) and phosphatidylcholine (PC)
are more abundant in the outer leaflet\cite{Alberts2002} leading
to a permanent dipole density for the membrane, similarly as in
monolayers. In fact, it is well established that a difference in
electric potential across a membrane due to dipolar potentials is
quite significant, between 100 mV and 400 mV, and that this
potential difference is extremely important for a variety of
biological processes \cite{Brockman1994}.

What is the effect of domains contained in a matrix with a
different dipole density? In a series of papers Andelman et al.
\cite{Andelman1985,Andelman1987} and McConnell and collaborators
\cite{Keller1987,McConnell1988,Lee1993,Lee1994,McConnell1996} have
provided theoretical and experimental evidence that dipolar
domains in monolayers have a profound effect in the phase diagram,
as the electrostatic free energy cost of assembling such domains
limits its size, preventing the monolayer from reaching complete
phase separation. In their original calculation, McConnell and
collaborators did not consider several effects that are
potentially relevant, such as a finite ionic strength, the
inclusion of image charges and the polarizability of the dipoles.
Those effects were considered by Andelman et al.
\cite{Andelman1987} for geometries consisting of stripes.
Furthermore, in view of the importance of lipid domains in
membranes, it is of great interest to analyze how dipole density
inhomogeneities may affect the phase diagram \cite{Smorodin2001},
similarly as it is the case for simple monolayers at the air-water
interface. In this paper, we generalize the calculation by
McConnell and collaborators for circular domains to include finite
ionic strength, image charges and the polarizability of the
dipoles and further extend the results to dipolar domains in
bilayers.

\begin{figure}
\includegraphics[width=8cm]{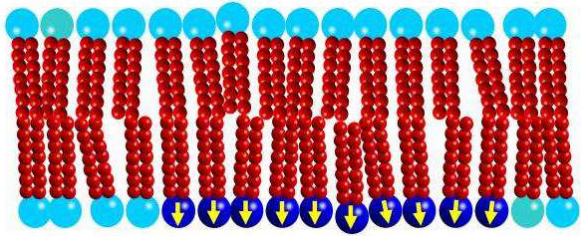}
\caption{An asymmetric bilayer with a domain with a dipole
density.} \label{fig:Bilayer}
\end{figure}

The phase diagram of lipid mixtures is a result of the microscopic
characteristics of the lipids involved such as hydrocarbon chain
and length as well as degree of saturation, possibility of forming
hydrogen bonds, etc.. It is expected that the effect of these
interactions is parameterized on longer scales by quantities such
as a line tension $\gamma$ between coexisting domains, a
spontaneous curvature $c_0$ of the different phospholipid species,
the membrane surface tension $\tau$ or its bending rigidity
$\kappa$ (see for example \cite{Gozdz2001,Harden2005} and
references therein). The incorporation of the microscopic
permanent dipoles of the lipids in these long wavelength
descriptions is subtle. As an example, let us consider a bilayer
containing a circular domain with non-zero dipolar density within
a matrix of an hypothetical dipole free lipid, as shown in
Fig.~\ref{fig:Bilayer}. The electrostatic free energy cost of such
domain can be estimated in analogy with the monolayer case
\cite{Keller1987,McConnell1988}
\begin{equation}\label{eq:Mc_log}
    F \approx R (a+b\log R) \ ,
\end{equation}
where $R$ is the size of the domain. If the coefficient $b$ were
vanishingly small, the net electrostatic cost of the dipoles grows
with the perimeter of the domain and may be lumped into a
``renormalized'' line tension contribution. For the case shown in
Fig.~\ref{fig:Bilayer}, however, the coefficient $b$ is not
negligible, and the free energy contribution of the permanent
dipoles must be considered explicitly. In some other situations,
however, the coefficient $b$ may vanish. It seems plausible that
if the domain in Fig.~\ref{fig:Bilayer} was opposed by an
identical domain of the same lipid, the membrane would not exhibit
a permanent dipole moment and the electrostatic contribution would
consist of a short-ranged interaction that might be incorporated
into the line tension. Furthermore, ionic strength, by weakening
the electrostatic interactions, can also influence domain sizes.
In this paper we present a detailed quantitative analysis for
these questions.

In all previous treatments of dipolar domains (such as
\cite{McConnell1988} and all subsequent work), the free energy has
been computed by directly summing over dipoles. In this paper, we
will consider two oppositely charged sheets and obtain the free
energy of the dipoles as the first non-trivial coefficient in the
limit of small sheet separation. Obtaining the dipolar result in
this way is particularly convenient for implementing the boundary
conditions that should be met by dipole densities within layers
with different dielectric constants.

The organization of the paper is as follows. In
Sect.~\ref{SECT__FreeEnergy} we derive an explicit form for the
free energy of the system. A summary of the exact solution for the
electrostatic free energy of dipolar domains is provided in
Sect.~\ref{SECT__Exact}. The details of the solution are
mathematically dense so we defer the complete derivation to
appendix~\ref{app_Elect}. In Sect.~\ref{SECT__Analysis} we analyze
the theoretical implications for monolayers and bilayers domains.
We conclude our paper with a discussion on some of the
consequences of our results in Sect.~\ref{SECT__Conclusions}.

\section{Electrostatic Free Energy}\label{SECT__FreeEnergy}

We consider a system of three slabs with different dielectric
constants and Debye lengths. Two disks of opposite surface charges
parallel to each other are located on the boundaries of the middle
slab, as shown in Fig.~\ref{fig:Cases}. The distance $D$ between
the two oppositely charged disks is assumed smaller than any other
length in the problem and therefore the two disks represent a
dipolar density. In the dilute electrolyte regime, the free energy
of the system consists of an ideal and an electrostatic term,
\cite{Safran1994}
\begin{eqnarray}\label{eq:Free_Energy}
  F &=&k_BT \sum_{a} \int d^3{\bf r} n_a({\bf r})(\log(v n_a({\bf
   r}))-1)+\nonumber\\&+&
\frac{1}{8\pi} \int d^3 {\bf r} \vec{E}\cdot \vec{D} \ ,
\end{eqnarray}
where $n_a({\bf r})=n_a^B \exp(-\frac{q_a e \Phi ({\bf r})}{k_B
T})$ is the number density of ions of type $a$, ionic valence
$q_a$ and bulk concentration $n_a^B$. It is convenient to express
the free energy in a different form. For the ideal part we write
\begin{eqnarray}\label{eq:Entropy}
  &&\sum_{a} \int d^3{\bf r} n_a({\bf r})(\log(v n_a({\bf
   r}))-1)\nonumber \\
   &=&-\sum_{a} \frac{q_a e}{k_B T}\int d^3{\bf r} n_a({\bf r})\Phi ({\bf
   r})+\sum_{a} N_a(\log(n_a^Bv)-1)\nonumber\\
   &=&-\sum_{a} \frac{q_a e}{k_B T}\int d^3{\bf r} n_a({\bf r})\Phi ({\bf
   r})+\sum_{a}N_a(\log(v\frac{N_a}{V})-1)-\nonumber\\
   &-&\sum_{a} n_a^B \int d^3{\bf r}(\exp(-\frac{q_a e \Phi ({\bf r})}{k_B
T})-1) \ ,
\end{eqnarray}
where the last term follows from the identity $N_a=\int d^3{\bf r}
n_a^B\exp(-\frac{q_a e \Phi ({\bf r})}{k_B T})$, where $N_a$ is
the number of ions of type $a$ within the solution. The
electrostatic term is written as
\begin{eqnarray}\label{eq:Electrostatic}
&&\frac{1}{8\pi} \int d^3 {\bf r} \vec{E}\cdot
\vec{D}=-\frac{1}{8\pi} \int d^3 {\bf r} \nabla \Phi ({\bf
r})\cdot \vec{D}
\\
&=&\frac{1}{2}\int d^3 {\bf r} e \Phi ({\bf r})\sum_{a} q_a
n_a({\bf r})+\frac{1}{2}\int d^3 {\bf r} e \Phi ({\bf
r})\sigma({\bf r})\nonumber
\end{eqnarray}
The previous expressions are general for an interface in contact
with a dilute ionic solution. Dipolar interactions are weak so we
further assume the ``Debye-Huckel'' approximation $\frac{e \Phi
({\bf r})}{k_B T}<<1$. The ion number density becomes
\begin{equation}\label{eq:linear_ap} n_a({\bf r})=n_a^B(1-q_a\frac{e \Phi
({\bf r})}{k_B T}+\frac{1}{2} \frac{q_a^2 e^2 (\Phi ({\bf
r}))^2}{(k_B T)^2}+..)  .
\end{equation}
Upon using the charge neutrality condition $\sum_{a}q_{a} n_{a}^B
=0$, it follows
\begin{eqnarray}\label{eq:identitities}
\sum_{a} q_a n_a({\bf r})&=&-(\sum_{a} q_a^2 n_a^B )\frac{e \Phi
({\bf r})}{k_B T}+\cdots\\\nonumber \sum_{a} q_a \int d^3 {\bf r}
n_a({\bf r})\frac{e\Phi ({\bf r})}{k_B T}&=&-\sum_{a}  q_a^2 n_a^B
\int d^3 {\bf r}(\frac{e\Phi ({\bf r})}{k_B T})^2 .
\end{eqnarray}
Inserting Eq.~\ref{eq:Entropy}, Eq.~\ref{eq:Electrostatic},
Eq.~\ref{eq:linear_ap} and Eq.~\ref{eq:identitities} into
Eq.~\ref{eq:Free_Energy} the free energy is
\begin{equation}\label{eq:Free_simplified}
    F=\frac{1}{2}\int d^3 {\bf r} \Phi ({\bf
r})\sigma({\bf r})+k_B T \sum_{a} N_a (\ln(\frac{N_a}{V}v)-1).
\end{equation}
The first term is the electrostatic free energy of the charged
disks. It is remarkable that both the entropic and the
electrostatic terms involving the ions explicitly cancel out. The
second term is the ideal free energy of the ion species in
solution. The subsequent analysis will be performed within the
canonical ensemble where the total number of particles of the
different species is fixed, so the ideal free energy is constant
and will be ignored herein. For uniformly charged interfaces, the
free energy can be further simplified to
\begin{equation}\label{eq:Free_simpler}
    F=\pi |\sigma| \int_0^{\infty} d\rho \rho
    (\Phi(\rho,-\frac{D}{2})-\Phi(\rho,\frac{D}{2})) .
\end{equation}
A complete evaluation of the free energy requires the
determination of the electric potential, which follows from the
Poisson-Boltzmann (PB) equation
\begin{equation}\label{eq:PB}
    \nabla^2 \Phi ({\bf r})=\frac{1}{\lambda_D^2}\Phi ({\bf r})
\end{equation}
where $\lambda_D$ is the Debye length. In regions where no ions
are present ($\lambda_D=\infty$) and the previous equation reduces
to the standard Poisson equation. In this paper, we consider
problems with cylindrical symmetry. The solution to the PB
equation is
\begin{equation}\label{eq:PB_sol_cyl}
\Phi (\rho,z)=a(k)\exp(\pm k z)J_0(\rho(k^2-1/\lambda_D^2)^{1/2})
\ ,
\end{equation}
where $J_0(x)$ is the 0th order Bessel function and $k$ indexes
different solutions.

The electric potential is
\begin{widetext}
\begin{equation}\label{eq:Potential_general}
  \Phi (\rho,z)=\left\{
  \begin{array}{l c}
              \int^{\infty}_0 dk J_0(k \rho) \frac{\exp(-(k^2+1/(\lambda_D^a)^2)^{1/2}z)}{(k^2+1/(\lambda_D^a)^2)^{1/2}} a_1(k) & z>D/2 \\
              \int^{\infty}_0 dk J_0(k \rho) ( \frac{\exp((k^2+1/(\lambda_D^b)^2)^{1/2}z)}{(k^2+1/(\lambda_D^b)^2)^{1/2}}a_2(k)+\frac{\exp(-(k^2+1/(\lambda_D^b)^2)^{1/2}z)}{(k^2+1/(\lambda_D^b)^2)^{1/2}}a_3(k)) & -D/2<z<D/2\\
              \int^{\infty}_0 dk J_0(k \rho)  \frac{\exp((k^2+1/(\lambda_D^c)^2)^{1/2}z)}{(k^2+1/(\lambda_D^c)^2)^{1/2}} a_4(k) &
              z< -D/2
  \end{array} \right. \ ,
\end{equation}\
\end{widetext}
The condition that the electric field must vanish at infinity
($E_{z}(\infty)=E_{z}(-\infty)=0$) has been already implemented.
The values of $\lambda_D^{a,b,c}$ are the Debye lengths in each
region. We further assume a different dielectric constant
$\varepsilon^{a,b,c}$ for each region.

The coefficients $\{a_i(k)\}_{i=1\cdots 4}$ are determined from
the boundary conditions
\begin{eqnarray}\label{eq:boundary}
\Phi (\rho,(D/2)^{+})&=&\Phi (\rho,(D/2)^{-})\nonumber\\
\Phi (\rho,(-D/2)^{+})&=&\Phi (\rho,(-D/2)^{-})\nonumber\\
\varepsilon_a E_z(\rho,(D/2)^{+})-\varepsilon_b
E_z(\rho,(D/2)^{-})&=&-4\pi |\sigma| \Theta(R-\rho) \nonumber\\
\varepsilon_b E_z(\rho,(-D/2)^{+})-\varepsilon_c
E_z(\rho,(-D/2)^{-})&=&4\pi |\sigma| \Theta(R-\rho) \nonumber\\
&& \ ,
\end{eqnarray}
where $E_z$ is the component of the electric field along the
z-axis, which is assumed to be perpendicular to the plane defined
by the disks. The boundary conditions defined by
Eq.~\ref{eq:boundary} are easily implemented by recalling the
following identity for Bessel functions
\begin{equation}\label{eq:Identity_Bessel}
    \Theta(R-\rho)= \int^{\infty}_0 dk J_0(k \rho)J_1(kR) \ ,
\end{equation}
where $\Theta(x)$ is Heaviside-$\Theta$ function defined by
$\Theta(x)=1$ if $x>0$ and $\Theta(x)=0$ for $x<0$.

For further reference, we quote the free energy of two infinite
disks of uniform and opposite charges filled with a medium of
dielectric constant $\varepsilon_I$. The free energy of this
system is equivalent to the elementary formula of an infinite
capacitor
\begin{equation}\label{eq:Elementary_Cap}
    F_C=\frac{2\pi^2}{\varepsilon_I}|\sigma|^2R^2D.
\end{equation}
We also will need the non-ideal free energy for the Guoy-Chapman
diffuse layer of a weakly charged system $\lambda_D/\lambda_G<<1$,
where $\lambda_D$ is the Debye length and $\lambda_G$ is the
Guoy-Chapman length. The free energy is given by
\begin{equation}\label{eq:Diffuse_Layer}
    F_{GC}=\frac{2\pi^2}{\varepsilon_w}|\sigma|^2R^2 \lambda_D .
\end{equation}
We note that upon replacing $\varepsilon_I$ by the dielectric
constant of the solvent $\varepsilon_w$ and the finite size $D$ by
the Debye length $\lambda_D$, the ``capacitor''formula
Eq.~\ref{eq:Elementary_Cap} is identical with the previous
equation. With the explicit form of the potential
Eq.~\ref{eq:Potential_general} introduced into
Eq.~\ref{eq:Free_simplified}, the free energy of the system is
\begin{equation}\label{eq:Lang_Helm}
    F=\frac{2\pi^2}{\varepsilon_I}|\sigma|^2R^2D{\cal
    F^I}(\frac{D}{R},\frac{R}{\lambda_D}) \ ,
\end{equation}
where ${\cal F^I}(x,y)$ is a function specific for each situation,
which according to Eq.~\ref{eq:Elementary_Cap} satisfies ${\cal
F^I}(0,0)=1$. In the following we provide an explicit calculation
for the function ${\cal F^I}(x,y)$ in different cases.

\section{Exact results for the free energy in several cases}\label{SECT__Exact}

\begin{figure}
\includegraphics[width=8cm]{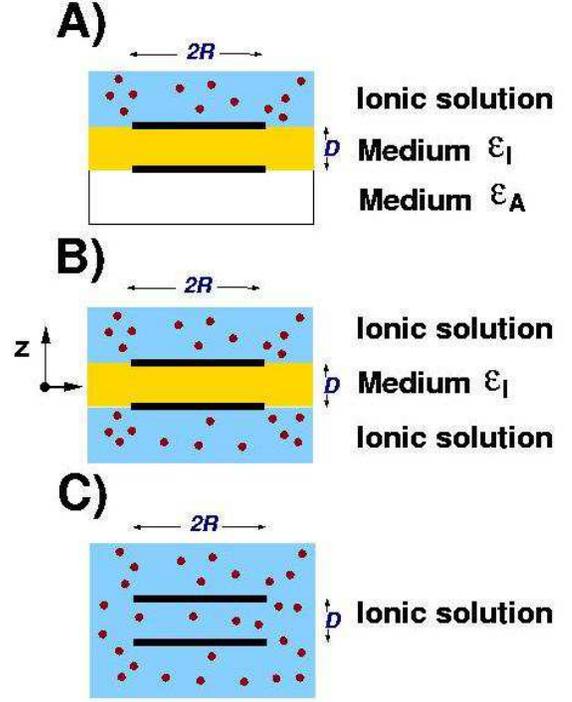}
\caption{Two domains of uniform and opposite charges in contact
with: A) one side with a salty solution, a medium in between and a
different medium in the outside. B) Both sides in contact with a
salty solution and a medium in between. C) submerged in a salty
solution.} \label{fig:Cases}
\end{figure}

We provide an explicit solution for the following three problems
shown in Fig.~\ref{fig:Cases}. In all cases, two infinitely thin
disks of finite radius $R$ and uniform but opposite charges are
separated by a distance $D$, and it is assumed that $R>>D$. In
Problem A) the two disks are on the boundary of a medium of
dielectric constant $\varepsilon_I$, with only one domain in
contact with a salty solution with solvent of dielectric constant
$\varepsilon_w$ (water) and the other domain in contact with a
medium of dielectric $\varepsilon_A$. In problem B) both domains
are in contact with a salty solution with solvent of dielectric
$\varepsilon_w$, while the interior contains a polarizable medium
of dielectric constant $\varepsilon_I$. In Problem C) the two
domains are submerged in a solution with solvent of dielectric
$\varepsilon_w$.

The free energy of these three systems can be computed exactly.
The most relevant results are summarized below. Explicit formulas
and complete details of the derivation are provided in
appendix~\ref{app_Elect}.

\subsection{Finite circular dipole domain in a medium of dielectric $\varepsilon_I$ with one
side in contact with a salty solution and the other with a medium
of dielectric $\varepsilon_A$ }\label{sub_SECT__ProblemA}

The free energy for this system is given by the general expression
\begin{eqnarray}\label{eq:summary_A}
     F^A&=&\frac{2\pi^2}{\varepsilon_I}|\sigma|^2R^2D\left(1- \right.\\\nonumber
     &-&\frac{D}{\pi
     R}\left. \{
     f_1(\frac{R}{\lambda_D})\ln(R/D)+f_2(\frac{R}{\lambda_D})\}\right)
\end{eqnarray}
where $f_1(y)$ and $f_2(y)$ are analytic functions of $y$. In the
limit $y\rightarrow 0$ ($R/\lambda_D<<1$, low screening limit) and
$y\rightarrow \infty$ ($R/\lambda_D>>1$, high screening limit)
both $f_1$ and $f_2$ can be computed analytically. The results for
$f_1$ are
\begin{eqnarray}\label{eq:f_prob_A}
f_1(0)&=&2\frac{\varepsilon_w\varepsilon_A}{\varepsilon_I(\varepsilon_w+\varepsilon_A)}\nonumber\\
f_1(\infty)&=&2\frac{\varepsilon_A}{\varepsilon_I}
\end{eqnarray}
Explicit formulas for $f_2$ and further details of the calculation
are provided in appendix~\ref{app_SECT__ProblemA}.

\subsection{Finite circular dipole domain within a
medium of dielectric $\varepsilon_I$ both sides in contact with an
aqueous solution}\label{Sub_SECT__ProblemB}

The free energy adopts different forms. In the low screening limit
it reduces to a similar expression as in the previous case,
\begin{equation}\label{eq:summary_B_low}
F^B=\frac{2 \pi^2}{\varepsilon_I}|\sigma|^2 R^2[1-\frac{D}{\pi
R}(f_1\ln(R/D)+f_2)]
\end{equation}
where $f_1=\frac{\varepsilon_w}{\varepsilon_I}$ and the explicit
expression for $f_2$ is quoted in Eq.~\ref{eq:f_Low_salt}.

In the high screening limit, there are two situations that need to
be considered, $R>>\lambda_D >>D$ and $R>>D>>\lambda_D$. For
$R>>\lambda_D >>D$ it is found
\begin{equation}\label{eq:summaryB_high_1}
    F^B=2 \frac{2 \pi^2}{\varepsilon_I}|\sigma|^2 R^2 D(1+{\cal
    O}(D/\lambda_D)) .
\end{equation}
This result is derived for a simplified problem discussed below.
In the limit $R>>\lambda_D>>D$ it follows
\begin{equation}\label{eq:summaryB_high}
    F^B=2 \frac{2 \pi^2}{\varepsilon_w}|\sigma|^2 R^2 \lambda_D .
\end{equation}
Explicit formulas and detailed calculations are provided in
appendix~\ref{app_SECT__ProblemB}.

\subsection{Two oppositely charged disks in solution}\label{sub_SECT__ProblemC}

The low screening limit has the same form as in Problem A. In the
high screening limit, and for $D>>\lambda_D$, the system consists
of two infinitely thin disks with diffuse layers on both sides
interacting through a screened Coulomb potential,
\begin{equation}\label{eq:SECT__ProblemC}
    F^C=\frac{2
    \pi^2}{\varepsilon_w}|\sigma|^2R^2\lambda_D(1-\exp{(-D/\lambda_D)})
    .
\end{equation}
In the high screening limit, but for $R>>\lambda_D>>D$, the
leading term of the free energy is the ``capacitor'' term and is
given by
\begin{equation}\label{eq:Inter_screening}
    F^C=\frac{2
    \pi^2}{\varepsilon_I}|\sigma|^2R^2D(1-\frac{D}{2 \lambda_D}) \
    .
\end{equation}
The first correction is independent of the radius $R$ of the
domain. The details of the calculation are provided in
appendix~\ref{app_SECT__ProblemC}.

\section{Implications for monolayers and lipid domains}\label{SECT__Analysis}

\begin{figure}
\includegraphics[width=8 cm]{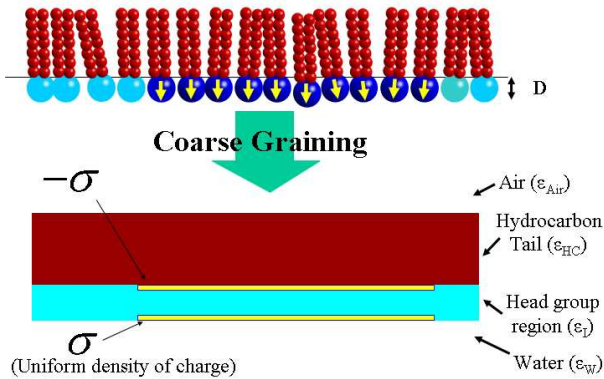}
\caption{A Langmuir monolayer with a domain with a finite dipole
density. The monolayer head group is represented as a medium with
dielectric constant $\varepsilon_{I}$ within two opposite charge
planes with uniform surface charges, one side in contact with an
aqueous solution and the other with a medium of dielectric
constant $\varepsilon_{HC}$ (the hydrocarbon tail).}
\label{fig:Monolayer}
\end{figure}

\subsection{Monolayers}\label{Sub_SECT_Monolayer}

We consider a monolayer with dielectric constant $\varepsilon_I$
in between water $\varepsilon_w\approx80$ and air
$\varepsilon_{Air}\approx1$. The monolayer consists of a minority
phase of $n_s$ domains with radius $R$ and dipolar density
$\mu\equiv \sigma D$ contained within a matrix of a majority phase
with zero dipolar density, as shown in Fig.~\ref{fig:Monolayer}.
The area spanned by the minority phase is $A_s\equiv n_s \pi R^2$.
The free energy is constructed from the coarse-graining suggested
in Fig.~\ref{fig:Monolayer}, which consists of approximating the
head group region as a medium of dielectric constant
$\varepsilon_I$ and the hydrocarbon tail as an infinite medium of
dielectric constant $\varepsilon_{HC}$. From the results in
Subsect.~\ref{sub_SECT__ProblemA} it is
\begin{equation}\label{eq:Free_monolayer}
    F_{tot}=(2\pi \gamma R + F^A)\frac{A_s}{\pi R^2} ,
\end{equation}
where $\gamma$ is the line tension between the two phases. The
free energy explicitly depends on the domain radius $R$, and is
minimized under the constraints of both constant dipolar density
$\mu$ and minority phase area $A_s$. The equilibrium radius of the
domains $R_{eq}$ is given from
\begin{equation}\label{eq:Radius_domain}
    R_{eq}=D \exp(1-\frac{f_2}{f_1}) \exp(\frac{\varepsilon_I \gamma}{f_1
    \mu^2})\approx D \exp(\frac{\varepsilon_I \gamma}{f_1
    \mu^2}).
\end{equation}
The above expression is only exact in the limits $y\rightarrow 0$
(low screening) and $y\rightarrow \infty$ (high screening), as
otherwise terms involving derivatives of $f(y)$ are present. From
Eq.~\ref{eq:f_prob_A}, as a result of
$\varepsilon_w>>\varepsilon_A=\varepsilon_{HC}$, it is found
$f_1(0)\approx f_1(\infty)$. It also follows from
Eq.~\ref{eq:f_0_Lang} that $f_2(0)\approx f_2(\infty)$. Therefore,
for both the low and high screening limits, the domain radius is
the same and is given by
\begin{equation}\label{eq:Radius_limiting}
  R_{eq}\approx D\exp(\frac{\varepsilon_I^2\gamma}{2\varepsilon_{HC}\mu^2}).
\end{equation}
The term in the exponent is only half of the value originally
obtained by McConnell \cite{Keller1987,McConnell1988} and it is
multiplied by the factor $\varepsilon_I^2/\varepsilon_{HC}$. For
the case $\varepsilon=1$, it agrees, upto the 1/2 factor, with the
result for stripes obtained in \cite{Andelman1987}. The origin of
the $1/2$ factor is the discontinuity in the dielectric constant
between air and water, while the presence of the different
dielectric constants is due to the finite polarizability of the
dipoles and the surrounding medium.

The same formula applies for a more general situation where the
matrix has a non-zero dipole density $\mu_1$ and the domains a
dipolar density $\mu_2$. In that case, the quantity appearing in
the exponent is the dipolar density difference
$\mu=|\mu_2-\mu_1|$.

\subsection{Bilayers}

We now consider a bilayer with a circular domain, where the
dipolar density of the outer leaflet ($\mu_1$) and the inner
leaflet ($\mu_2$) are different. Similarly as in the case for
monolayers, the dipolar densities are to be interpreted as the
excess dipolar density from the matrix reference state. The domain
has radius $R$ and the vertical separation between the two
leaflets is $L$, as shown in fig.~\ref{fig:Bilayer_CG}. The
bilayer is surrounded by water ($\varepsilon_w\approx 80$). As
shown in appendix~\ref{app_Domains_across}, the electrostatic free
energy cost is of the form
\begin{equation}\label{eq:free_diff_in_mu}
F_{ele} \sim -(\mu_1-\mu_2)^2R\ln(\frac{8R}{L}) +\gamma^\prime
2\pi R ,
\end{equation}
where terms proportional to $R^2$ have been ignored as they are
not relevant for this argument. For a symmetric domain
($\mu_1=\mu_2$) there is no logarithmic contribution and the
electrostatic free energy is of the same form as the line tension.
The above result has been derived without including the effects of
ionic strength or different dielectric constants, but we assume it
true under these more general conditions.

\begin{figure}
\includegraphics[width=8 cm]{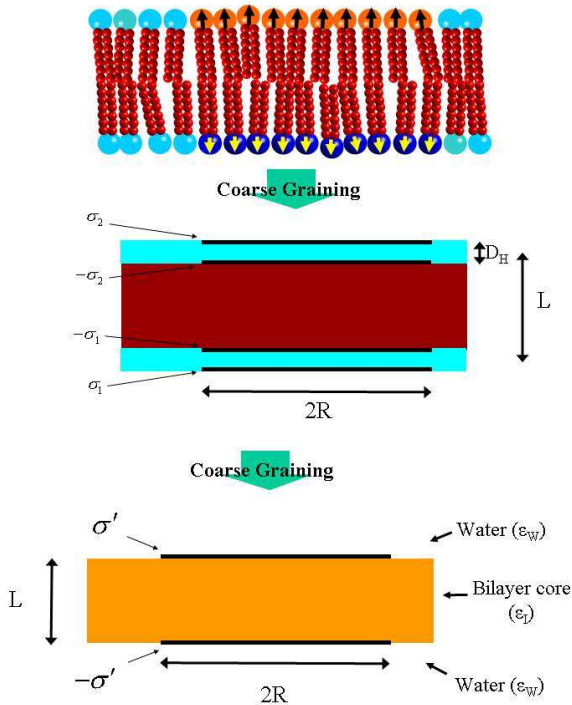}
\caption{Dipolar domains in a bilayer. In a first approximation,
the two leaflets of the bilayer are modelled as two monolayers, an
inner leaflet monolayer with dipolar density $\mu_2\equiv \sigma_2
D$ and an outer leaflet monolayer with $\mu_1\equiv\sigma_1 D$,
following the same steps as in Fig.~\ref{fig:Monolayer}. At longer
distances, the net dipolar moment of the two monolayers can be
represented as if were originated by two oppositely charged disks
with charge density $\sigma^{\prime}=(\sigma_1-\sigma_2)D/L$, with
an effective dielectric constant $\varepsilon_I$ taking into
account both hydrocarbon tail and head groups polarizabilities.}
\label{fig:Bilayer_CG}
\end{figure}

If the membrane contains an intrinsic asymmetry between the inner
and outer leaflets (for definiteness, we assume $\mu_1\ > \mu_2$),
there is a net dipolar moment. We consider domains much larger
than the vertical separation of the two leaflets ($R>>L$), and
construct the free energy by first coarse-graining each separate
leaflet of the bilayer following the same steps as in the
monolayer case. The bilayer has a net dipolar moment, which we
model as if it were produced by two disks of uniform but opposite
surface charge density $\sigma^{\prime}$, where
$\sigma^{\prime}=(\sigma_1-\sigma_2)D_H/L$ and $D_H$ is the width
of the head group region. This argument should not be understood
as implying that the coarse-graining of an asymmetric bilayer
should lead to equal and opposite charge distributions, but rather
that this configuration leads to the same dipolar moment. It is
expected that higher order moments provide sub-leading
corrections, which nevertheless may be incorporated as any other
short-range interactions, if necessary. The dielectric constant
$\varepsilon_I$ is an effective value that includes contributions
both from the hydrocarbon tails and the head group and therefore
it should be expected that $\varepsilon_I\sim 4$. The different
steps in the coarse graining process are illustrated in
Fig.~\ref{fig:Bilayer_CG}. Under these assumptions, the problem of
an asymmetric domain is reduced to the problem discussed in
Subsect.~\ref{Sub_SECT__ProblemB}. In the low screening limit, the
radius of $n_s$ domains is computed as for the monolayer case, and
is given by
\begin{equation}\label{Radius_Bilayer}
R_{eq} \approx L
\exp(\frac{\varepsilon_I^2\gamma}{\varepsilon_W(\mu_1-\mu_2)^2}) \
,
\end{equation}
where $\varepsilon_I$ is the effective dielectric constant
including contributions from the hydrocarbons and the
polarizabilities of the dipoles. This formula differs from the
analogous result for monolayers in the factor
$\varepsilon_w/2\varepsilon_{HC} \approx 10$ in the exponent,
which implies that if a given dipole density and line tension in a
monolayer leads to an equilibrium radius $R_{eq}\approx 10 \mu M$,
the same dipolar density and line tension in a bilayer results in
much smaller domains, of radius comparable to the molecular scale
(of the order of $L$). Alternatively, the previous formula shows
that the condition for large (micron sized) domains to exist is
given by
\begin{equation}\label{eq:dipole_moment}
    \frac{\varepsilon_I^2\gamma}{\varepsilon_w \mu^2} \approx 8
     \rightarrow \Delta p \approx \varepsilon_I A_0\sqrt{\frac{k_B
    T}{8 \varepsilon_w L^{\prime}}} \ ,
\end{equation}
where $p$ is the permanent moment of the phospholipid, $A_0$ is
the molecular area and we assumed that the line tension is
$\gamma\approx \frac{k_B T}{L^{\prime}}$. With $A_0 \approx 60$
\AA$^2$, $L^{\prime}\approx 10$ \AA, the formula yields $\Delta
p<< 0.1 \varepsilon_I$ (D). If we assume $\varepsilon_I \approx 4$
then $\Delta p<< 0.4 $ D. That is, a net dipolar moment difference
of the order of $0.4$ D or less is needed for large (micron sized)
dipolar domains to be observed. Dipolar moments for phospholipids
are much larger, of the order of $15$ D \cite{Gawrisch1992}, which
implies that domains with significant asymmetry will be sub-micron
in size.

In the high screening limit, the results are qualitatively
different. When the Debye length becomes of the order of the
domain radius, the electrostatic interactions become short-ranged
and the free energy, which is given in Eq.~\ref{eq:summaryB_high},
does not show the logarithmic contribution present in
Eq.~\ref{eq:free_diff_in_mu}. In that case, the thermodynamic
stable state of the $n_s$ domains is completely dominated by line
tension and the free energy is minimized by a single circular
domain (assuming that the resulting line tension is positive) of
the minority phase within a matrix of the majority phase. A
summary of the different results is shown in
Fig.~\ref{fig:Phse_diagram}.

\section{Discussion and conclusions}\label{SECT__Conclusions}

\begin{figure}
\includegraphics[width=8 cm]{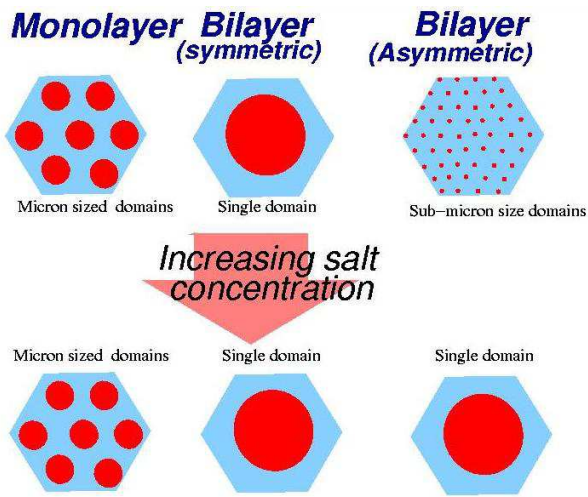}
\caption{Schematic representation of the different systems studied
upon variations in ionic strength.} \label{fig:Phse_diagram}
\end{figure}

We presented a detailed calculation showing how the competition
between line tension and electrostatic dipolar interactions
determines the size of lipid domains, thus extending previous
results by McConnell and collaborators
\cite{Keller1987,McConnell1988} by including finite ion strength,
variations in the dielectric constant and the polarizability of
the dipoles and further extending the results to account for
dipolar domains in lipid membranes. The main conclusions are
summarized in Fig.~\ref{fig:Phse_diagram}; for lipid monolayers at
the air-water interface it is found that the size of the lipid
domains is independent of the ionic strength, but is sensitive to
the discontinuities in dielectric constant and the polarizability
of the dipoles. In lipid vesicles, it has been shown that for
symmetric domains, where the net dipole density in the outer and
inner leaflet is the same, the electrostatic contribution to the
free energy can be absorbed into the line tension coefficient. For
asymmetric domains, which have a net dipolar density, it has been
shown that the large dielectric constant of water is responsible
for a dramatic decrease of the size of the domains as compared
with the size obtained for a monolayers with the same line tension
and dipole density. It has also been shown that small variations
in dipolar density (2$\%$ or more from the equilibrium dipole
density of a zwitterionic phospholipid) are enough to limit domain
sizes to sub-micron scales. Opposite to the situation with
monolayers, dipolar domains in bilayers are sensitive to solution
ionic strength. At small Debye lengths, of the order of the domain
radius at zero ionic strength, dipolar electrostatic effects
gradually become negligible and the system is dominated by line
tension, resulting in a single large circular minority domain if
the line tension is positive. Our study has been focused on the
limiting cases of high and low ionic strength, but exact analytic
expressions have been provided for the free energy and the
electric potential. Problems that have not been discussed in this
paper, such as values for domain radius at intermediate ionic
strength, ionic distributions or potential differences across the
interface also follow from these expressions.

Our results show the subtle role played by the intrinsic dipoles
in lipids membranes. At high ionic strength, the effect of dipoles
is short-ranged, but at low ionic strength dipolar interactions
are only negligible for symmetric domains, and in any other
situation, dipolar interactions must be included explicitly,
either in continuum models \cite{Gozdz2001,Harden2005} or in
numerical simulations \cite{Stevens2004}.

Recent experimental and theoretical studies have studied the
interactions of circular and planar dipolar domains
\cite{Khattari2002} and interactions within two domains
\cite{Wurlitzer2002}. As compared with previous theoretical
studies \cite{Smorodin2001} of domain shape transitions in lipid
bilayers, our study includes the effects of a net dipolar moment
and the dielectric constant of water. More importantly, it shows
that at high ionic strength the effect of dipolar interactions is
short-ranged. Other theoretical work \cite{Netz1999} has discussed
interfaces within the framework of the Debye-Huckel approximation,
the same approximation used in this study, but has been focused on
effective interactions between charges, so these results are
somewhat complementary to the ones presented here.

A number of situations have not been considered in this paper. Our
study has been restricted to circular domains, while other shapes,
such as stripes, are also of interest \cite{Keller1987,Seul1995}.
We further assumed a planar membrane, thus neglecting the effects
of Gaussian curvature, which has been shown that, at least for
solid domains, may lead to significant effects
\cite{Travesset2005,Travesset2006,Bowick2006}. The results
presented have ignored domain-domain interaction and its possible
superstructures. This problem is discussed in \cite{Seul1995},
where it is shown that triangular lattices of circular domains are
usually found, and a discussion is presented on the transition
from circular to other domains and other phases such as stripes
\cite{Keller1999}.

Problems where dipolar-like domains, similar to the ones
considered in this paper, have been discussed in other contexts.
The free energy describing the phase transition between a Fermi
liquid and an insulating Wigner crystal phase in a clean two
dimensional electron can be mapped to a dipolar interaction and a
line tension term. After a detailed analysis, the authors in
\cite{Spivak2003} showed that a direct first order phase
transition cannot occur, and instead, it was shown that the
transition should proceed as a sequence of intermediate
``microemulsion''phases, consisting of domains of one phase within
a matrix of the other phase. Our results are in complete agreement
with \cite{Spivak2003}, but show that a direct first order
transition can be obtained as a function of the Debye length, if
the electrostatic interactions are screened. Previous MD
simulations \cite{Reichhardt2003a,Reichhardt2003b} have shown that
transitions from triangular superlattices to striped domains are a
general feature resulting from increasing the strength of a
short-ranged attraction force with respect to a long range
repulsion. Upon identifying the short-range attraction with the
line tension and the long range repulsion with the dipolar
interactions, these results are qualitatively similar to the
problems studied in this paper. It seems therefore that
``microemulsion''phases are a common feature of systems near a
phase transition consisting of a long range repulsion and short a
range attraction. In fact, this has been shown rigorously for
general 2D Coulomb systems \cite{Jamei2005}. Another interesting
example is provided by heteroepitaxial growth of 2D islands in
solid surfaces, where the elastic free energy cost of assembling
an island also contains a logarithmic term (completely analogous
with the dipolar case studied in this paper) and competes against
the line tension. In \cite{Liu2001}, it was shown that the
addition of domain-domain interactions in the form of dipolar
interactions leads to the formation of triangular superlattices.
These results show that interactions among finite circular domains
are responsible for the formation of superlattice structures of
circular domains, whose size can be computed from balancing
elastic (or dipolar) free energy against line tension, and provide
yet another example illustrating the expected general results of
phase separations in systems with long range interactions.

It is quite interesting to compare the exact solutions discussed
in Sect.~\ref{SECT__Exact} with the corrections to a finite size
capacitor, as discussed Landau \ \& \ Lifshitz \cite{Landau2004}.
Although the charge distribution in the plates of a finite size
capacitor is not uniform, as we assume in our calculations, the
coefficient of the logarithmic term $f_1$ in the electrostatic
energy for $\varepsilon_I=\varepsilon_A=\varepsilon_W$,
Eq.~\ref{eq:f_0_Lang} is identical with the analogous formula
quoted in Landau \ \& \ Lifshitz \cite{Landau2004}.

This paper has shown the subtle role played by dipolar moments. We
showed that the zwitterionic nature of the phospholipids leads to
new effects both in bilayers and monolayers. It is our expectation
that the results presented will be useful in other areas.

{\bf acknowledgements}

I thank A. Grosberg and J. Israelachvili for proposing this
problem and for many discussions. I also acknowledge discussions
with C. Lorenz and D. Vaknin. I acknowledge the Benasque Center
for Sciences, where this work was started. This work is supported
by NSF grant DMR-0426597 and partially supported by DOE through
the Ames lab under contract no. W-7405-Eng-82.

\appendix

\section{Detailed derivation of the expressions for the free energy}\label{app_Elect}

\subsection{Finite circular dipole domain in a medium of dielectric $\varepsilon_I$ with one
side in contact with a salty solution and the other with a medium
of dielectric $\varepsilon_A$}\label{app_SECT__ProblemA}

This situation corresponds to case A) in Fig.~\ref{fig:Cases}.
There are three dielectric constants, the medium in between the
disks ($\varepsilon^b=\varepsilon_I$), the medium outside the
disks ($\varepsilon^c=\varepsilon_A$) and the dielectric of the
solvent $\varepsilon^a=\varepsilon_w$ (water). The electric
potential in the different region follows from
Eq.~\ref{eq:PB_sol_cyl} with both
$\lambda_D^b=\lambda_D^c=\infty$.

The coefficients $\{a_i(k)\}_{i=1\cdot4}$ are determined from the
boundary conditions Eq.~\ref{eq:boundary} with the help of
Eq.~\ref{eq:Identity_Bessel}. The explicit result for the
$a_4,a_3$ coefficients are
\begin{widetext}
\begin{eqnarray}\label{eq:Lang_coeff}
    a_1(k)&=&\frac{-(\sinh(kD/2)+\frac{\varepsilon_A}{\varepsilon_I}\cosh(kD/2))\sinh(kD/2)
    \exp((k^2+1/\lambda_D^2)^{1/2}D/2)}{\frac{\cosh(kD/2)}{\varepsilon_I}(\frac{\varepsilon_w}{k}+
    \frac{\varepsilon_A}{(k^2+1/\lambda_D)^{1/2}})+\sinh(kD/2)(\frac{\varepsilon_A\varepsilon_w}{k\varepsilon_I^2}+
    \frac{1}{(k^2+1/\lambda_D)^{1/2}})}\frac{8\pi|\sigma| R}{k
    \varepsilon_I} J_1(kR)\nonumber\\\nonumber
    a_4(k)&=&\frac{(\frac{k \sinh(kD/2)}{(k^2+1/\lambda_D^2)^{1/2}}+\frac{\varepsilon_w}{\varepsilon_I}\cosh(kD/2))\sinh(kD/2)
    \exp((k^2+1/\lambda_D^2)^{1/2}D/2)}{\frac{\cosh(kD/2)}{\varepsilon_I}(\frac{\varepsilon_w}{k}+
    \frac{\varepsilon_A}{(k^2+1/\lambda_D)^{1/2}})+\sinh(kD/2)(\frac{\varepsilon_A\varepsilon_w}{k\varepsilon_I^2}+
    \frac{1}{(k^2+1/\lambda_D)^{1/2}})}\frac{8\pi|\sigma| R}{k
    \varepsilon_I} J_1(kR)
    \\&&
\end{eqnarray}
\end{widetext}
The explicit form for free energy follows from
Eq.~\ref{eq:Free_simpler}
\begin{equation}\label{eq:Lang_Helm_app}
    F=\frac{2\pi^2}{\varepsilon_I}|\sigma|^2R^2D{\cal
    F^A}(\frac{D}{R},\frac{R}{\lambda_D}) \ ,
\end{equation}
where ${\cal F^A}(x,y)$ is given by
\begin{widetext}
\begin{equation}\label{eq:def_F}
    {\cal F^A}(x,y)=4\int^{\infty}_0
    \frac{dk}{k^2}J^2_1(k)\frac{\sinh(kx/2)}{x}\frac{\cosh(kx/2)(\frac{a}{k}+\frac{b}{(k^2+y^2)^{1/2}})+
    \sinh(kx)\frac{2}{(k^2+y^2)^{1/2}}}{\cosh(kx/2)(\frac{a}{k}+\frac{b}{(k^2+y^2)^{1/2}})+
    \sinh(kx)(\frac{ab}{k}+\frac{1}{(k^2+y^2)^{1/2}})} \ .
\end{equation}
\end{widetext}
In order to alleviate the notation, we have introduced new
variables $a=\varepsilon_w/\varepsilon_I$ and
$b=\varepsilon_A/\varepsilon_I$. In the strict $R>>D$ limit, the
free energy corresponds to two infinite circular plates, as the
function ${\cal F^A}$ satisfies
\begin{equation}\label{eq:Lang_Limit_F}
    \lim_{x\rightarrow0}  {\cal F^A}(x,y) = 1 \ ,
\end{equation}
irrespectively of the value for $y$. The first correction ${\cal
O}(D/R)$ to the free energy is of the form
\begin{eqnarray}\label{eq:Lang_smallx}
     F&=&\frac{2\pi^2}{\varepsilon_I}|\sigma|^2R^2D(1-\frac{D}{\pi
     R}\{
     f_1(\frac{R}{\lambda_D})\ln(R/D)+f_2(\frac{R}{\lambda_D})\})\nonumber
     \\&& \ ,
\end{eqnarray}
where $f_1(y)$ and $f_2(y)$ are two analytic functions, whose
limiting expressions in the low and high screening limits is
provided below. The general derivation of the expression
Eq.~\ref{eq:Lang_smallx} follows from the same steps as the
formula derived below for the restricted case of the low screening
limit.

\subsubsection{The low screening limit ($R << \lambda_D$)}

The function ${\cal F^A}$ must be considered in the limit
$y\rightarrow 0$,
\begin{widetext}
\begin{eqnarray}\label{eq:def_F_y_zero}
    {\cal F^A}(x,0)&=&\frac{4}{x} \int^{\infty}_0
    \frac{dk}{k^2}J^2_1(k)\sinh(kx/2)\frac{(a+b)\cosh(kx/2)+
    2\sinh(kx)}{(a+b)\cosh(kx)+(ab+1)
    \sinh(kx))}\\\nonumber
    &=&\frac{2}{x} \int^{\infty}_0
    \frac{dk}{k^2}J^2_1(k)(1-\exp{(-kx)})\frac{1+
    \frac{(2-a-b)}{2(a+b)}(1-\exp{(-kx)})}{1+\frac{(ab+1-a-b)}{2(a+b)}(1-\exp{(-2kx)})}.
\end{eqnarray}
\end{widetext}
This function is of the general form Eq.~\ref{app_GenericF}, where
$A=\frac{2-a-b}{2(a+b)}$ and $B=\frac{ab+1-a-b}{2(a+b)}$.  The
general expression for the asymptotic behavior follows from the
result in Eq.~\ref{app_expansionx},
\begin{eqnarray}\label{eq:f_0_Lang}
   f_1(0)&=&2\frac{ab}{a+b}=2\frac{\varepsilon_w\varepsilon_A}{\varepsilon_I(\varepsilon_w+\varepsilon_A)}\nonumber\\
   f_2(0)&=&2\frac{ab}{a+b}\ln(8/(e^{1/2}))+\frac{(2-a-b)(2a+2b)}{(a+b)^2}-\nonumber\\
   &-&\frac{2(a+b)}{ab+1+a+b}\left({\cal
   G}_1(\frac{ab+1-a-b}{ab+1+a+b})-\right.\nonumber\\
   &-&\left.\frac{4-2a-2b}{a+b}{\cal
   G}_2(\frac{ab+1-a-b}{ab+1+a+b})\right)
\end{eqnarray}
where the functions ${\cal G}_i$ are defined in
Eq.~\ref{app_G_function}. If both salt-free mediums have the same
dielectric constant $\varepsilon_I=\varepsilon_A$, then $b=1$ and
the expression for $f_2$ simplifies to
\begin{equation}\label{eq:f2_0_b_1}
    f_2(0)=\ln(8/e^{1/2})-\frac{1-a}{2(1+a)}\ln(4/e) .
\end{equation}

In \cite{Riviere1995} the coefficient $f_1(0)$ for a dipole domain
with no polarization $\varepsilon_I=1$ was quoted, and the result
is in agreement with Eq.~\ref{eq:f_0_Lang}. McConnell and
collaborators \cite{Keller1987,McConnell1988} have computed the
free energy in the limit
$\varepsilon_w=\varepsilon_A=\varepsilon_I=1$ by integrating the
individual contributions of single dipoles and introducing a
short-distance cut-off $\delta$ to avoid a self-energy
singularity. Their result is
\begin{eqnarray}\label{eq:f2_0_a_1}
    f_1(0)&=&1 \\\nonumber
    f_2(0)&=&\ln(e^2\delta/(4D)) \ ,
\end{eqnarray}
which is identical with Eq.~\ref{eq:f_0_Lang} if the cut-off is
taken as $\delta=D/(2 e^{3/2})$.

\subsubsection{The high screening limit $R>>\lambda_D$}

We now consider the limit where the Debye screening is much
shorter than the radius of the dipolar domain. We therefore need
to consider the limiting case $y \rightarrow \infty$ in the
scaling function Eq.~\ref{eq:def_F}. The resulting ${\cal F^A}$ is
\begin{widetext}
\begin{eqnarray}\label{eq:F_y_infty_Lang}
    {\cal F^A}(x,\infty)&=&\frac{2}{x} \int^{\infty}_0
    \frac{dk}{k^2}J^2_1(k)(1-\exp{(-kx)})\frac{1
    -\frac{1}{2}(1-\exp{(-kx)})}{1+\frac{(b-1)}{2}(1-\exp{(-2kx)})}
    \ ,
\end{eqnarray}
\end{widetext}
which is of the general form Eq.~\ref{app_GenericF}, with $A=-1/2$
and $B=(b-1)/2$. The expressions for $f_1(\infty)$ and
$f_2(\infty)$ coefficients follow as
\begin{eqnarray}\label{eq:f_y_infty_Lang}
    f_1(\infty)&=&2b=2 \frac{\varepsilon_A}{\varepsilon_I} \\\nonumber
    f_2(\infty)&=&2b\ln(8/e^{3/2})-\frac{2}{b+1}\left({\cal G}_1(\frac{b-1}{b+1})+\right.\\\nonumber
    &+&\left.2{\cal G}_2(\frac{b-1}{b+1})\right) \ .
\end{eqnarray}
Both $f_1(\infty)$ and $f_2(\infty)$ are independent of
$\varepsilon_w$. In the high screening limit, the electric field
does not penetrate in the water, and therefore no reference to the
dielectric properties of the water should be expected.

\subsection{Finite circular dipole domain within a
medium of dielectric $\varepsilon_I$ both sides in contact with an
aqueous solution}\label{app_SECT__ProblemB}

This situation differs from the previous case in that the aqueous
solution is in contact with both sides of the domain. The
potential is screened everywhere except for the thin layer in
between the two disks, where the dielectric constant is
$\varepsilon_I$. The electric potential follows from
Eq.~\ref{eq:Potential_general}, with
$\lambda_D^a=\lambda_D^c=\lambda_D$ and $\lambda_D^b=\infty$.

Upon imposing the boundary conditions Eq.~\ref{eq:boundary} the
coefficients satisfy $a_1(k)=-a_4(k)$ and $a_2(k)=-a_3(k)$, with
\begin{widetext}
\begin{eqnarray}\label{eq:Helm_coeff}
    a_1(k)&=&-(k^2+1/\lambda_D^2)^{1/2}\exp{((k^2+1/\lambda_D^2)^{1/2}D/2)}
    \frac{4\pi|\sigma|R
    J_1(kR)\sinh(kD/2)}{(k^2+1/\lambda_D^2)^{1/2}\varepsilon_w \sinh(kD/2)+k\varepsilon_I\cosh(kD/2)}
    \nonumber\\
    a_2(k)&=&-\frac{2\pi|\sigma|kR J_1(kR)}{(k^2+1/\lambda_D^2)^{1/2}\varepsilon_w\sinh(kD/2)+k\varepsilon_I\cosh(kD/2)}
\end{eqnarray}
\end{widetext}
and the free energy is
\begin{equation}\label{eq:Helm_Raft}
    F=\frac{2 \pi^2}{\varepsilon_I}|\sigma|^2R^2D{\cal
    F^B}(\frac{D}{R},\frac{R}{\lambda_D}) \ ,
\end{equation}
where ${\cal F^B}$ is defined from
\begin{widetext}
\begin{equation}\label{eq:R_def}
    {\cal F^B}(x,y)=\frac{8}{x}\int^{\infty}_0 dk
    J^2_1(k)\frac{\sinh(kx/2)}{k(k^2+y^2)^{1/2}}\frac{\cosh(kx/2)+\frac{\varepsilon_I
    k}{\varepsilon_w(k^2+y^2)^{1/2}}\sinh(kx/2)}{(\frac{\varepsilon_w}{\varepsilon_I}+\frac{k^2}{(k^2+y^2)})\sinh(kx)+\frac{2
    k}{(k^2+y^2)^{1/2}}\cosh(kx)} \ .
\end{equation}
\end{widetext}
In the low screening limit $R<<\lambda_D$, this function becomes
of the general form Eq.~\ref{app_GenericF} with
$A=\frac{1}{2}(\frac{\varepsilon_I}{\varepsilon_w}-1)$ and
$B=\frac{1}{4}(\frac{\varepsilon_I}{\varepsilon_w}+\frac{\varepsilon_I}{\varepsilon_w}-2)$,
thus the free energy becomes
\begin{equation}\label{eq:Helm_LowSalt_Raft}
F=\frac{2 \pi^2}{\varepsilon_I}|\sigma|^2 R^2(1-\frac{D}{\pi
R}\left\{f_1\ln(R/D)+f_2\right\}) \ ,
\end{equation}
where the coefficients are given from
\begin{eqnarray}\label{eq:f_Low_salt}
f_1&=&\frac{\varepsilon_w}{\varepsilon_I} \\\nonumber
f_2&=&\frac{\varepsilon_w}{\varepsilon_I}\ln(8/e^{1/2})+(\frac{\varepsilon_I}{\varepsilon_w}-1)(
\frac{\varepsilon_I}{\varepsilon_w}+\frac{\varepsilon_w}{\varepsilon_I})-\\\nonumber
&-&\frac{4}{2+\frac{\varepsilon_I}{\varepsilon_w}+\frac{\varepsilon_w}{\varepsilon_I}}\left({\cal
G}_1(\frac{\frac{\varepsilon_I}{\varepsilon_w}+\frac{\varepsilon_w}{\varepsilon_I}-2}
{\frac{\varepsilon_I}{\varepsilon_w}+\frac{\varepsilon_w}{\varepsilon_I}+2})-\right.\\\nonumber
&-&\left.2(\frac{\varepsilon_I}{\varepsilon_w}-1) {\cal
G}_2(\frac{\frac{\varepsilon_I}{\varepsilon_w}+\frac{\varepsilon_w}{\varepsilon_I}-2}
{\frac{\varepsilon_I}{\varepsilon_w}+\frac{\varepsilon_w}{\varepsilon_I}+2})\right)
\end{eqnarray}
These expressions are actually identical with
Eq.~\ref{eq:f_0_Lang} with
$a=b=\frac{\varepsilon_w}{\varepsilon_I}$.

The high screening limit ($\lambda_D <<R$) corresponds to
$y\rightarrow\infty$. In this limit,  ${\cal F}^B$ takes a
particularly simple form
\begin{equation}\label{eq:HighSalt_Raft}
{\cal F}^B(x,\infty)=\frac{4}{xy}\int^{\infty}_0 dk
\frac{J^2_1(k)\varepsilon_I}{k\varepsilon_w}=\frac{2\varepsilon_I}{xy\varepsilon_w}
\ ,
\end{equation}
leading to the free energy
\begin{equation}\label{eq:Helm_HighSalt_Raft}
    F=2 \frac{2 \pi^2}{\varepsilon_w}|\sigma|^2 R^2 \lambda_D .
\end{equation}
This result is significantly different from the low screening
limit, as the free energy is independent of the separation between
the two disks $D$, which has been replaced by the Debye length and
the dielectric constant is no longer the dielectric constant of
the membrane but that of the water. This expression is twice the
free energy of the diffuse layer Eq.~\ref{eq:Diffuse_Layer}, and
implies that the system minimizes the free energy by effectively
decoupling the two charged disks into two diffuse layers,
resulting in a zero electric field within the bilayer. This result
also follows from noticing that the ``Guoy-Chapman'' free energy
Eq.~\ref{eq:Helm_HighSalt_Raft} is lower that the ``capacitor''
formula Eq.~\ref{eq:Elementary_Cap} whenever
\begin{equation}\label{eq:cond_Raft}
    \frac{D}{\varepsilon_I}> \frac{2\lambda_D}{\varepsilon_w} \ ,
\end{equation}
which is always satisfied for a sufficiently small $\lambda_D$.
There is another regime defined by $R>>\lambda_D$ and
$\lambda_D>>D$ where the ``capacitor'' formula is still the
leading term, but its corrections are ${\cal O}(D/\lambda_D)$.
This regime is discussed in detail in the next problem.

\subsection{Two oppositely charged disks in solution}\label{app_SECT__ProblemC}

This situation is similar to the previous case, except that now
both the solvent and the ions can be found in between the two
plates. The electric potential follows from
Eq.~\ref{eq:Potential_general} with
$\lambda_D^a=\lambda_D^b=\lambda_D^c$. The boundary conditions are
given by the usual Eq.~\ref{eq:boundary}, but only one dielectric
constant appears ($\varepsilon_I=\varepsilon_A=\varepsilon_w$).
The $a_i(k)$ coefficients satisfy $a_3(k)=-a_2(k)$ and
$a_1(k)=-a_4(k)$, with
\begin{eqnarray}\label{eq:PC_coeff}
    a_1(k)&=&-\frac{4\pi}{\varepsilon_w} |\sigma|R
    J_1(kR)\sinh((k^2+1/\lambda_D^2)^{1/2}D/2)
    \nonumber\\
    a_2(k)&=&-\frac{2\pi}{\varepsilon_w} |\sigma|R
    J_1(kR)\exp(-(k^2+1/\lambda_D^2)^{1/2}D/2)
    \nonumber\\
    &&
\end{eqnarray}
The free energy is
\begin{equation}\label{eq:Helm_Coll}
    F=\frac{2 \pi^2}{\varepsilon_w}|\sigma|^2R^2D{\cal
    F^C}(\frac{D}{R},\frac{R}{\lambda_D}) \ ,
\end{equation}
where ${\cal F^C}$ is defined by
\begin{equation}\label{eq:Q_def}
    {\cal F^C}(x,y)=\frac{2}{x}\int^{\infty}_0 dk
    J^2_1(k)\frac{1-\exp{(-(k^2+y^2)^{1/2}x})}{k(k^2+y^2)^{1/2}}
\end{equation}
The low screening limit defined by $\lambda_D>>R$ ($y \rightarrow
0$) has already been considered as it is identical as the one
considered in Sect.~\ref{app_SECT__ProblemA} ( ${\cal
F^C}(x,0)={\cal F^A}(x,0)$), whose limit is provided by
Eq.~\ref{eq:f_0_Lang}.

In the high screening limit ($R>>\lambda_D$), we consider two
situations corresponding to either $D>>\lambda_D$ or
$D<<\lambda_D$.  In both cases the ${\cal F^C}$ function
Eq.~\ref{eq:Q_def} needs to be evaluated for $y>>1$, but with the
additional assumptions that either $xy>>1$ or $xy<<1$. For $xy>>1$
it is readily obtained
\begin{equation}\label{eq:Q_screening}
 {\cal F^C}(x,y)=\frac{1}{xy}(1-\exp{(-xy)}) \ .
\end{equation}
and the resulting free energy is
\begin{equation}\label{eq:Helm_screening}
    F=\frac{2
    \pi^2}{\varepsilon_w}|\sigma|^2R^2\lambda_D(1-\exp{(-D/\lambda_D)})
\end{equation}
The first term is the diffuse layer free energy of two finite,
infinitely thin disks surrounded by a diffuse layer of counterions
on both sides. In this case, each disk has a contribution that is
exactly 1/2 of Eq.~\ref{eq:Diffuse_Layer}. The second term
represents the Coulomb attraction of the two disk-charges, which
is screened by the electrolytes.

The limit $xy<<1$ is more complex, and it is obtained from the
following function
\begin{eqnarray}\label{eq:app_Q_split}
{\cal Q_J}(x,y)&=&\int^{y}_0 dk
    J^2_1(k)\frac{1-\exp{(-(k^2+y^2)^{1/2}x})}{k}\nonumber\\
    &+&
    \int^{\infty}_y dk
    J^2_1(k)\frac{1-\exp{(-(k^2+y^2)^{1/2}x})}{k}\nonumber\\
    && \ ,
\end{eqnarray}
which is related as $\frac{\partial{(x{\cal F^C}})}{{\partial
x}}=2 {\cal Q_J}$ (see Eq.~\ref{eq:Q_def}). The first integral is
expanded as
\begin{eqnarray}\label{eq:app_Q_sp_zero}
&&\int^{y}_0 dk
    J^2_1(k)\frac{1-\exp{(-(k^2+y^2)^{1/2}x})}{k}
    \\
    &=&\int^{y}_0 \frac{dk}{k}
    J^2_1(k)-xy\int^{y}_0 \frac{dk}{k}
    J^2_1(k)(1+k^2/y^2)^{1/2}\nonumber\\\nonumber
    &=&\int^{\infty}_0 \frac{dk}{k}
    J^2_1(k)-\int^{\infty}_y \frac{dk}{k}
    J^2_1(k)+xy\int^{\infty}_0\frac{dk}{k}
    J^2_1(k)\\\nonumber
    &=&\frac{1-xy}{2}-\int^{\infty}_y \frac{dk}{k}
    J^2_1(k)+{\cal O}((xy)^2).
\end{eqnarray}
The second integral is computed from
\begin{eqnarray}\label{eq:app_Q_sp_infty}
&&\int^{\infty}_y dk
    J^2_1(k)\frac{1-\exp{(-(k^2+y^2)^{1/2}x})}{k}
    \\\nonumber
    &=&\frac{2}{\pi}\int^{y}_0 \frac{dk}{k^2}
    (1-\sin(2k))\exp(-(k^2+y^2)^{1/2}x)\\\nonumber
    &=&\frac{2}{\pi y}\int^{\infty}_1
    \frac{du}{u^2}(1-\sin(2yu))\exp(-xy(1+u^2)^{1/2})\\\nonumber
    &=&\frac{2}{\pi y}(1+{\cal O}(xy))\ .
\end{eqnarray}
where the asymptotic expansion of the Bessel functions for large
values of the argument have been used. Combining equations
Eq.~\ref{eq:app_Q_sp_zero} and Eq.~\ref{eq:app_Q_sp_infty}, and
noticing that the terms of order $1/y$ cancel, it follows that
\begin{equation}\label{eq:app_Q_J}
{\cal Q_J}(x,y)=\frac{1-xy}{2}+{\cal O}((xy)^2,(xy)/y) \ ,
\end{equation}
which leads to the small x-expansion for ${\cal F^C}$,
\begin{equation}\label{eq:app_Q_low_xy}
{\cal F^C}(x,y)=\frac{2}{x}(\frac{x}{2}-\frac{x^2y}{4}+\cdots) \ ,
\end{equation}
or
\begin{equation}\label{eq:Q_Interm_regime}
    {\cal F^C}(x,y)=1-\frac{xy}{2} \ .
\end{equation}
The free energy in the regime defined by $D<<\lambda_D$ and
$R>>\lambda_D$ is therefore
\begin{equation}\label{eq:Inter_screening_app}
    F=\frac{2
    \pi^2}{\varepsilon_I}|\sigma|^2R^2D(1-\frac{D}{2 \lambda_D}) \ ,
\end{equation}
that is, the first correction is independent of the domain radius
$R$, consistent with the short-range nature of the screened
potential.

\section{Interaction of domains across a bilayer}\label{app_Domains_across}

In this appendix we consider two uniform dipole densities
separated a distance $L$, as shown in the middle picture of
Fig.~\ref{fig:Bilayer_CG}. In this situation we consider that both
dipole densities are surrounded by a media with the same
dielectric constant and no salt. The electric potential can be
obtained from the coefficients Eq.~\ref{eq:Lang_coeff} into
Eq.~\ref{eq:Potential_general} and using the superposition
principle. Introducing this potential into the free energy
Eq.~\ref{eq:Free_simpler} the following expression follows
(assuming L$>$D).
\begin{eqnarray}\label{app_Free_across}
    F&=&F_C^1+F_C^2-2\pi(\sigma_1-\sigma_2)^2D^2R\ln(\frac{8R}{L})+D^2 R {\cal H}(\frac{L}{D})\nonumber\\
    &&
\end{eqnarray}
where $F_C^i$ is the capacitor free energy
Eq.~\ref{eq:Elementary_Cap} for a surface charge $\sigma_i$ and
the function ${\cal H}$ is independent of $R$. If the dipole
densities are the same $\sigma_1=\sigma_2$, the interaction free
energy cost is linear in the radius R and can be incorporated into
the line tension.

\section{Expansion for ${\cal F}$}\label{app_calF}

In many of the situations we consider in this paper, we have to
find the expansion for small $x$ of a function
\begin{eqnarray}\label{app_GenericF}
    {\cal F}(x)&=&\int^{\infty}_0 \frac{dk}{k^2}
    J^2_1(k)(1-\exp{(-kx)})\times \nonumber\\
    &\times& \frac{1+A(1-\exp{(-kx)})}{1+B(1-\exp{(-2kx)})} \ ,
\end{eqnarray}
where $A,B$ are parameters that depend on the particular problem.
Upon expanding the denominator in powers of $\exp{(-kx)}$ and
using the integrals Eq.~\ref{Integ1} and Eq.~\ref{Integ2},
 the
following expression follows
\begin{eqnarray}\label{app_expansionx}
  {\cal F}(x) &=& \frac{x}{2}-\frac{x^2}{2\pi}(1+4B-2A)\ln(8/(e^{1/2}x))
  \\\nonumber
  &+&\frac{x^2}{2\pi}\frac{1}{1+B}({\cal G}_1(\frac{B}{1+B})-4A{\cal
G}_2(\frac{B}{1+B}))\\\nonumber
   &+&2 \frac{x^2}{\pi}A(1+2B)
\end{eqnarray}
where ${\cal G}_i$ are functions defined explicitly
\begin{eqnarray}\label{app_G_function}
{\cal G}_1(z)&=&\sum_{n=1}((2n+1)^2\ln(2n+1)-4n^2\ln(n))z^n
\nonumber\\
{\cal G}_2(z)&=&\sum_{n=1}((n+1)\log(n+1)-n\log(n))z^n \ ,
\end{eqnarray}
which are both convergent for $|z|<1$. It follows from
Eq.~\ref{app_expansionx} that if the coefficient in
Eq.~\ref{app_GenericF} satisfies $B>-1/2$, the argument of ${\cal
G}$ is $|z|<1$, thus ensuring the convergence of the results.

\section{Two relevant integrals}\label{app_use_int}

The following results are used throughout the calculation and are
quoted without further derivation. These integrals are calculated
to the first non-trivial order in an expansion in powers of its
argument $x$.

\begin{eqnarray}\label{Integ1}
\int^{\infty}_0 \frac{dk}{k^2}
J^2_1(k)(1-\exp{(-kx)})\exp(-nkx)&=&\nonumber\\\nonumber
=\frac{x}{2}-\frac{x^2}{2\pi}\left[(2n+1)\ln(8/(e^{1/2}x))-(n+1)^2\ln(n+1)\right.&+&\\
\left.+n^2\ln(n)\right] &&
\end{eqnarray}

\begin{eqnarray}\label{Integ2}
\int^{\infty}_0 \frac{dk}{k^2}
J^2_1(k)(1-\exp{(-kx)})^2\exp(-nkx)&=&\nonumber\\\nonumber
=\frac{x^2}{2\pi}\left[2\ln(8/(e^{1/2}x))-(n+2)^2\ln(n+2)+n^2\ln(n) \right]&&\\
 &&
\end{eqnarray}

%REFERENCE LIST

\end{document}